\pdfoutput=1
\documentclass[conference, 10pt]{IEEEtran}
\IEEEoverridecommandlockouts

\usepackage{color}
\usepackage{times,amsmath,epsfig}
\usepackage{amssymb}
\usepackage{subcaption}
\usepackage{cite}
\usepackage{url}
\usepackage[capitalise]{cleveref}
\usepackage[ruled,linesnumbered]{algorithm2e}
\usepackage{tikz, pgfplots}
\usepackage{moresize}
\usepackage{array}
\usepackage{amsthm}
\newtheorem{theorem}{Theorem}

\newtheorem{lemma}{Lemma}
\usepackage{booktabs}
\usepackage[table]{xcolor} 

\input{mysymbol.sty}

\pgfplotsset{compat=1.17}
\pgfplotstableset{col sep=semicolon}
\usepgfplotslibrary{fillbetween}
\tikzset{every mark/.append style={scale=1.5, solid}, font=\small}
\pgfplotsset{
    width=1.05\textwidth,
    height=5.5cm,
    legend style={
        font=\ssmall ,  
        inner xsep=1pt,
        inner ysep=1pt,
        nodes={inner sep=1pt}},
    legend cell align=left,
    every axis/.append style={line width=.5pt},
 	every axis plot/.append style={line width=1.5pt},
 	every axis y label/.append style={yshift=-4pt}
}
\usepgfplotslibrary{external}
\title{{Dirichlet Meets Horvitz and Thompson: Estimating Homophily in Large Networks via Sampling}}
\author{{Hamed Ajorlou$^*$,
        Gonzalo Mateos$^*$,
        and~Luana Ruiz$^\dagger$} \\
        $^*$Dept. of Electrical and Computer Engineering, University of Rochester, Rochester, NY\\
        $^\dagger$Dept. of Applied Mathematics and Statistics, Johns Hopkins University, Baltimore, MD \\
        
}

\begin{document}
%
\maketitle
\begin{abstract}
Assessing homophily in large-scale networks is central to understanding structural regularities in graphs, and thus inform the choice of models (such as graph neural networks) adopted to learn from network data. Evaluation of smoothness metrics requires access to the entire network topology and node features, which may be impractical in several large-scale, dynamic, resource-limited, or privacy-constrained settings. In this work, we propose a sampling-based framework to estimate homophily via the Dirichlet energy (Laplacian-based total variation) of graph signals, leveraging the Horvitz-Thompson (HT) estimator for unbiased inference from partial graph observations. The Dirichlet energy is a so-termed total (of squared nodal feature deviations) over graph edges; hence, estimable under general network sampling designs for which edge-inclusion probabilities can be analytically derived and used as weights in the proposed HT estimator. We establish that the Dirichlet energy can be consistently estimated from sampled graphs, and empirically study other heterophily measures as well. Experiments on several heterophilic benchmark datasets demonstrate the effectiveness of the proposed HT estimators in reliably capturing homophilic structure (or lack thereof) from sampled network measurements.

\end{abstract}

\section{Introduction}\label{s:introduction}

Homophily is a cardinal principle at the heart of several graph-based statistical learning tasks, including nearest-neighbor prediction, semi-supervised learning, and topology inference~\cite{kolaczyk_2009,luan2024heterophily,mateos2019topid}. Accordingly, assessing homophily characteristics (i.e., the extent to which edges preferentially connect nodes with similar attributes or labels) of  network datasets is central to understanding structural regularities in graphs, and, e.g., inform the choice of learning architectures such as graph neural networks (GNNs)~\cite{gama2020gnns}. In fact, a recent trend is to develop models for heterophilous data as well~\cite{luan2024heterophily, raghuvanshi2025heterophily,Liang2023GLRM}.

\vspace{2pt}
\noindent\textbf{Problem statement.} Consider a weighted and undirected graph $G = (\ccalV, \ccalE)$,  with  $n = |\ccalV|$  nodes, adjacency matrix \( \mathbf{A}=[A_{ij}] \in \mathbb{R}_{+}^{n \times n} \), and combinatorial graph Laplacian $\bbL=\textrm{diag}(\bbA \mathbf{1})-\bbA$. Let \( \mathbf{X}_n \in \mathbb{R}^{n\times f} \) be a graph signal matrix, where row vector $\bbx_i^\top$ collects the $f$ features at node $i\in\ccalV$. A workhorse homophily (i.e., signal smoothness) metric is the Dirichlet energy or Laplacian-based total variation defined as
\begin{equation}\label{eq:dirichlet}
    \textrm{TV}_G(\bbX_n) := \mathrm{trace}(\bbX_n^\top \bbL \bbX_n)= \sum_{(i,j) \in \ccalE} A_{ij} \|\bbx_i - \bbx_j\|^2,
\end{equation}
where smaller $\textrm{TV}_G(\bbX_n)$ values are indicative of homophilous feature characteristics in the data. The evaluation of $\textrm{TV}_G(\bbX_n)$ relies on the implicit assumption that the graph dataset is observed in its entirety; however, it is often the case that relational information is only acquired from a portion of the complex system of interest. Network \emph{sampling} is arguably the rule rather than the exception in statistical network analysis~\cite[Ch. 5]{kolaczyk_2009}, either due to uncontrollable factors behind the data acquisition process; or, by design in large-scale, dynamic, resource-limited, or privacy-constrained settings. 

Suppose that instead of fully observing $G$, we take measurements that effectively produce a sampled graph $G^* = (\ccalV^*, \ccalE^*)$ and signals $\bbX_{n^*}$, $n^*\leq n$. Consequently, it will be typically impossible to exactly recover $\textrm{TV}_G(\bbX_n)$ from the partial information in $G^*$. Instead, we study the \emph{problem} of developing useful homophily estimates of $\textrm{TV}_G(\bbX_n)$, say $\widehat{\textrm{TV}}$, from $G^*$ and $\bbX_{n^*}$ under various sampling designs~\cite{kolaczyk_2009,ahmed2014network,frank1980estimation}.

\vspace{2pt}
\noindent\textbf{Contributions in context.} Sampling has been widely studied in graph signal processing, but often with the objective of reconstructing bandlimited graph signals from limited nodal samples (yet $G$ is known in its entirety)~\cite{tanaka2020sampling}. Recent efforts have explored efficient training of GNNs from intentionally sampled homophilous graphs~\cite{li2024graph}. Our distinct goal is to estimate a global structural characteristic of a network (here homophily) in an unbiased fashion, and with quantifiable performance. 

In Section \ref{sec:testability}, we start by establishing a fundamental property of the estimation problem, namely that the Dirichlet energy is a so-termed \emph{testable} graph parameter under induced subgraph sampling. As introduced in~\cite{borgs2008convergentI}, testability of a network summary statistic is tantamount to the existence of an estimator satisfying a weak form of consistency as the sample size increases. The upshot is that, for testable parameters, one would expect that simple (e.g., plug-in $\widehat{\textrm{TV}}=\textrm{TV}_{G^*}(\bbX_{n^*})$) estimators will be accurate under induced subgraph sampling. For our technical arguments, we extend the seminal characterizations of testability based solely on convergence of graph sequences to graph limits (i.e., graphons)~\cite{borgs2008convergentI}, to also accommodate sequences of convergent graph signals~\cite{Ruiz_2021}. Although the testability result has merit to ensure the estimation task is feasible under induced subgraph sampling, other sampling designs could be of interest as well. And it is not uncommon for these to produce unequal probability sampling of nodes or edges, challenging the reliability of \emph{biased} plug-in estimators~\cite[Ch. 3]{kolaczyk_2017}. To bridge this gap, in Section \ref{sec:ht} we propose an homophily estimation framework by leveraging the Horvitz-Thompson (HT) estimator for unbiased inference from incomplete graph measurements. The Dirichlet energy  \eqref{eq:dirichlet} is a \emph{total} (of squared nodal feature deviations) over graph edges. Thus, estimable under general network sampling designs for which edge-inclusion probabilities can be analytically derived and used as weights in the proposed HT estimator. Interestingly, the variance of $\widehat{\textrm{TV}}$ can be readily estimated from the sample $G^*$ and $\bbX_{n^*}$.

We experimentally verify the unbiasedness of the HT estimator and examine how different sampling rates influence its variance (Section \ref{s:numerical_eval}). To gain further insight into the behavior of the estimator across a range of practical scenarios, we compare multiple network sampling designs and heterophily measures beyond \eqref{eq:dirichlet}. All in all, the main contributions of this work can be summarized as follows:
\begin{itemize}
    \item We establish that the Dirichlet energy is a testable graph parameter under induced subgraph sampling.
    
    \item We develop a novel HT estimator for unbiased estimation of homophily measures under general sampling designs.
    
    \item We conduct a comprehensive experimental evaluation using several heterophilic benchmark datasets. 
\end{itemize}

\section{Testability of the Dirichlet Energy}\label{sec:testability}

This section leverages results from the theory of graph limits to establish that the Dirichlet energy is a \emph{testable} graph (signal) parameter.
Following \cite[Def. 2.11]{borgs2008convergentI}, a graph parameter $\varphi$ is said to be testable if for every $\varepsilon>0$ there exists a sample size $n^*(\epsilon)$ such that for any graph $G$ with $n \ge n^*(\epsilon)$, an estimate $\widehat{\varphi}(G^{*})$ computed from a uniformly sampled induced subgraph $G^{*}$ satisfies
\begin{equation*}
\Pc{\, |\varphi(G)-\widehat{\varphi}(G^{*})| > \varepsilon }
\le \varepsilon .
\end{equation*}
Accordingly, testability of $\varphi$ implies the existence of an estimator $\widehat{\varphi}(G^{*})$ that satisfies a weak form of consistency.

A key result~\cite[Prop. 2.12(a)]{borgs2008convergentI} asserts that $\varphi$ is testable if and only if $\varphi(G_i)$ converges for every convergent graph sequence $G_1,G_2,\ldots$, which naturally leads to graphons and the theory of graph limits~\cite{Borgs2006Counting,borgs2008convergentI}. Thus, a suitable notion of continuity of $\varphi$ suffices. 
To establish the testability of the Dirichlet energy (a more general parameter than the graph signal-agnostic $\varphi(G)$ studied in~\cite{borgs2008convergentI}), we first recall the graph limit framework and explain how it naturally extends to continuous domain representations of graph signals~\cite{Ruiz_2021}.

A \emph{graphon} is a symmetric measurable function
$\bbW:[0,1]^2 \mapsto \reals$ that arises as the limit of a sequence of dense graphs
w.r.t. the cut metric.
The cut norm, which underlies the notion of continuity in graph limit theory,
is defined by
\[
    \|\bbW\|_{\square}
    := \sup_{S,T \subseteq [0,1]}
       \Bigl|\,
           \int\!\!\int_{S\times T} \bbW(u,v)\,du\,dv
       \Bigr|.
\]
It measures the maximum discrepancy of $\bbW$ over measurable cuts $S\times T$.
Similarly, graph signals on finite graphs admit continuous domain counterparts on the graphon, represented as measurable functions $\bbX:[0,1]\mapsto \reals^f$. Every finite graph–signal pair $(G,\bbX_n)$ admits a representation on a continuous domain via an associated step graphon and step signal.
To make this construction explicit, 
partition $[0,1]$ into intervals
\[
I_i=\Big[\tfrac{i-1}{n},\tfrac{i}{n}\Big),\qquad i=1,\ldots,n,
\]
and define the associated step graphon and step signal
\begin{align}
&\bbW_G(u,v)=\sum_{i=1}^{n}\sum_{j=1}^{n}
A_{ij}\,\ind{u\in I_i}\ind{v\in I_j}, \label{eq:step_graphon}\\
& \bbX_G(u)=\sum_{i=1}^{n}\bbx_i\,\ind{u\in I_i},\label{eq:step_signal}
\end{align}
where $\ind{\cdot}$ is an indicator. Thus, for any $u\in I_i$ and $v\in I_j$,
\[
\bbW_G(u,v)=A_{ij},
\qquad
\bbX_G(u)=\bbx_i,
\qquad
\bbX_G(v)=\bbx_j.
\]

In the limit of large graphs ($n\to\infty$), the step graphon and step signal converge to $(\bbW,\bbX)$~\cite{Ruiz_2021}. Now, the Dirichlet energy can be naturally extended to the graphon functional
\begin{align}\label{eq:functional}
\Phi(\bbW,\bbX)
= \iint_{[0,1]^2} \bbW(u,v)\,\|\bbX(u)-\bbX(v)\|^{2}\,du\,dv.
\end{align}
The following lemma shows that evaluating the continuous functional
defined in~\eqref{eq:functional} on a step graphon–signal pair
recovers the normalized Dirichlet energy of $\bbX_n$ in the corresponding finite graph $G$. Proofs are omitted due to lack of space, and will be reported in the extended journal version of this paper.
\begin{lemma}\label{lem:step_graphon_exact} 
Consider the step graphon--signal pair $(\bbW_G,\bbX_G)$ in \eqref{eq:step_graphon}-\eqref{eq:step_signal}. Then, the graphon functional $\Phi$ in \eqref{eq:functional} satisfies
\begin{align}\label{eq:graphon_discrete_exact}
\Phi(\bbW_G,\bbX_G) = &{}
\iint_{[0,1]^{2}}
\bbW_G(u,v)\,\|\bbX_G(u) - \bbX_G(v)\|^2\,du\,dv \nonumber\\
=
& {}\frac{1}{n^2}\sum_{(i,j) \in \ccalE}A_{ij}\,\|\bbx_i-\bbx_j\|^2\nonumber \\ 
= &{} \frac{1}{n^2}\mathrm{TV}_G(\bbX_n).
\end{align}
\end{lemma}

Based on the fundamental result of Borgs et al.~\cite[Proposition 5.10]{Borgs2006Counting} a graph parameter $\varphi$ is \emph{testable} if and only if it admits a graphon extension $\Phi:\mathcal{W} \mapsto \reals$ on the space of graphons $\mathcal{W}$ that is continuous w.r.t. the rectangle norm, equivalently the cut norm. Thus, the key steps to establish testability of the Dirichlet energy is to show that it admits a graphon extension $\Phi$ (cf. Lemma~\ref{lem:step_graphon_exact}) and that $\Phi$ is Lipschitz continuous w.r.t. the cut norm, as we show next.

\begin{theorem}\label{thm:dirichlet_testable}
The graphon functional $\Phi$ in~\eqref{eq:functional} is Lipschitz continuous under graphon--signal pair convergence.  
Specifically, if $\|\bbW_{G}-\bbW\|_{\square}\to 0$ and
$\|\bbX_{G}-\bbX\|_{2}\to 0$, then
\begin{equation}
\Phi(\bbW_{G},\bbX_{G}) \to \Phi(\bbW,\bbX), \quad n\to\infty.
\end{equation}
\end{theorem}

Because induced subgraph sampling for a large enough sample size $n^*$ yields sufficiently accurate approximations $G^*$ to $G$ in terms of cut distance~\cite[p. 47]{kolaczyk_2017}, then reliable homophily estimates are feasible for (even graph signal-dependent) continuous metrics such as \eqref{eq:dirichlet} and its associated graphon functional \eqref{eq:functional} as per Theorem \ref{thm:dirichlet_testable}. This naturally justifies the adoption of a simple plug-in estimator.

\section{Horvitz--Thompson Estimator for Homophily}\label{sec:ht}

To estimate homophily under \emph{general} sampling designs, we bring to bear ideas that (in network settings) can be traced back to the foundational work by O. Frank~\cite{frank2005sampling}; see also~\cite[Ch. 5]{kolaczyk_2009}. The challenge here is that network sampling often violates the standard assumption of a sample comprising i.i.d. observations from the population graph and, in fact, it often yields unequal probability sampling (see also Section \ref{ssec:designs}). A key requirement is that the network statistic of interest can be expressed as a total (or average) over sampled units, here unordered node pairs $\ccalV\times \ccalV$ or directly edges in $\ccalE$ [cf. \eqref{eq:dirichlet}].

Consider a random sampling design applied to $G$, which results in a sampled graph $G^* = (\ccalV^*, \ccalE^*)$ and signals $\bbX_{n^*}$, $n^*\leq n$. Let $\pi_{ij}:=\Pc{(i,j)\in\ccalE^*}$ be the \emph{inclusion probability} of edge $(i,j)$, meaning the probability that edge $(i,j)$ is sampled. Suppose that for the sampled edges $(i,j)\in\ccalE^*$ we observe (without error) the squared variation $V_{ij}:=A_{ij}\|\bbx_i-\bbx_j\|^2$.  
In this context, we propose a HT estimator~\cite{horvitz1952generalization} for the Dirichlet energy \eqref{eq:dirichlet}, namely
\begin{equation}\label{eq:ht_dirichlet}
    \widehat{\textrm{TV}}_{\textrm{HT}} := \sum_{(i,j) \in \ccalE^*} \frac{A_{ij}\|\bbx_i - \bbx_j\|^2}{\pi_{ij}}=\sum_{(i,j) \in \ccalE^*} \frac{V_{ij}}{\pi_{ij}},
\end{equation}
which is \emph{unbiased} for arbitrary sampling designs if \( \pi_{ij} > 0 \) for all edges. Indeed, using inclusion probabilities $\pi_{ij}^{-1}$ as weights is essential for unbiasedness since for the plug-in estimator $\widehat{\textrm{TV}}_{G^*}(\bbX_{n^*})$ one has
\begin{align}\label{eq:bias_plug_in}
    \E{\widehat{\textrm{TV}}_{G^*}(\bbX_{n^*})} ={}& \E{\sum_{(i,j) \in \ccalE^*} A_{ij}\|\bbx_i - \bbx_j\|^2}\nonumber\\
    ={}& \E{\sum_{(i,j) \in \ccalE} V_{ij}\ind{(i,j) \in \ccalE^*}}\nonumber\\
    ={}&\sum_{(i,j) \in \ccalE} V_{ij}\pi_{ij}.
\end{align}
From \eqref{eq:bias_plug_in}, $\E{\widehat{\textrm{TV}}_{\textrm{HT}}}=\textrm{TV}_G(\bbX_n)$ and unbiasedness follows. Notice that as it is customary in statistical sampling theory which relies on design-based inference, the only randomness is due to the sampling of nodes and edges. Graph data, in particular the $V_{ij}$, are asummed error free.


The variance of $\widehat{\textrm{TV}}_{\textrm{HT}}$ can be estimated from $G^*$ using 
\begin{equation}\label{eq:ht_dirichlet_var}
    \var{\widehat{\textrm{TV}}_{\textrm{HT}}}=\sum_{(i,j) \in \ccalE^*}\sum_{(k,l) \in \ccalE^*}V_{ij}V_{kl}\left(\frac{1}{\pi_{ij}\pi_{kl}}-\frac{1}{\pi_{ijkl}}\right),
\end{equation}
where now $\pi_{ijkl}:=\Pc{(i,j)\textrm{ and }(k,l)\in\ccalE^*}$ is the \emph{joint} inclusion probability of a pair of edges $(i,j)$ and $(k,l)$. Apparently, applicability of the HT framework hinges on three essential requirements: (i) the network statistic is expressible as a total over sampled units; (ii) the sampling design makes it feasible to observe the unit attribute of interest (here $V_{ij}$); and (iii) the (joint) inclusion probabilities can be calculated to form the estimator and evaluate its variance. Next, we outline canonical network sampling designs and comment on (iii).

\begin{figure*}[t]
    \centering
    \includegraphics[width=0.75\textwidth]{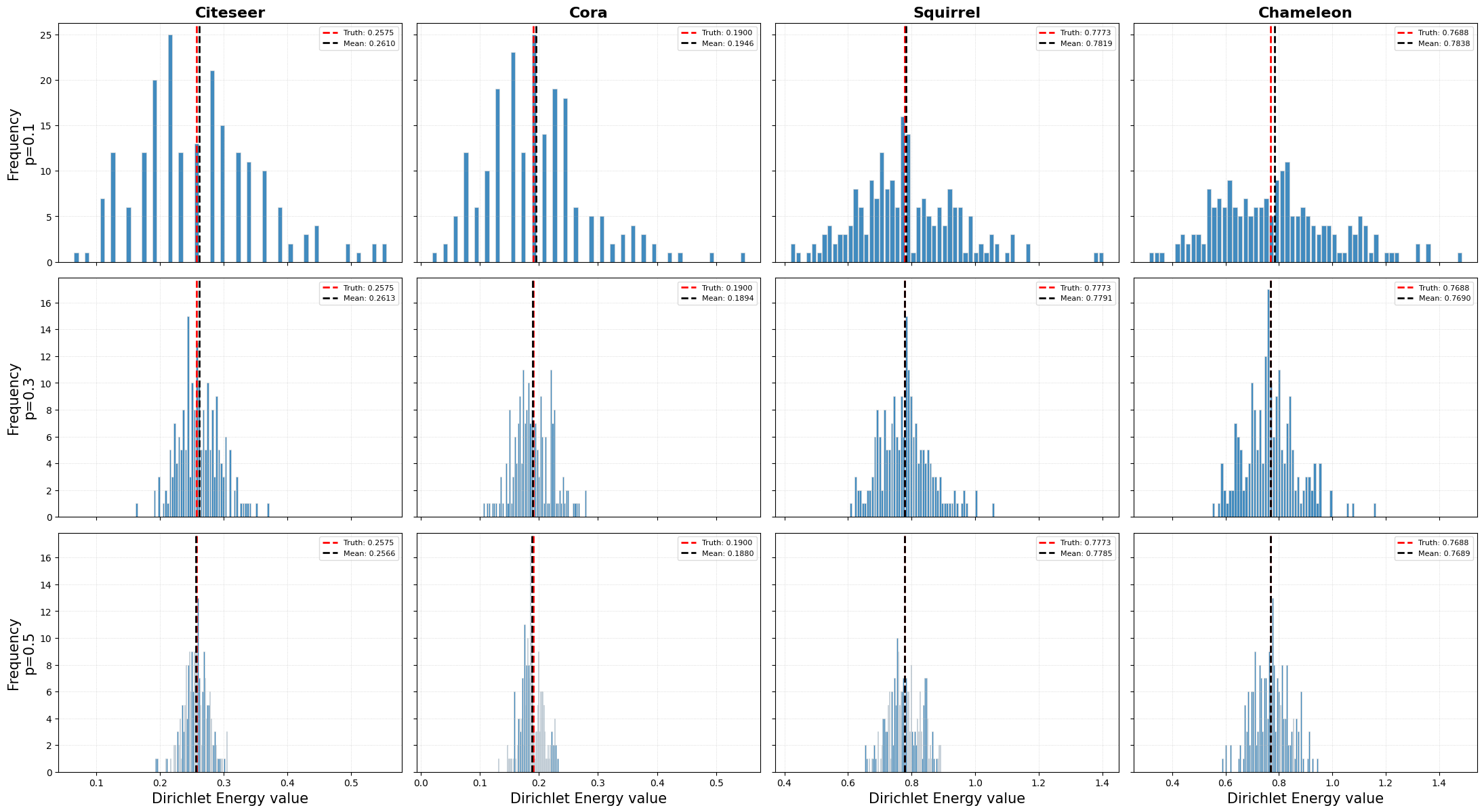}
    \caption{Dispersion analysis of $\widehat{\textrm{TV}}_{\textrm{HT}}$. Nodes are sampled independently with probability $p=\{0.1, 0.3, 0.5\}$ (BS), and edges are selected using induced subgraph sampling. Histogram of estimates over $T=200$ realizations, for varying $p$ and different datasets. Unbiasedness is well supported in all cases, while the estimator variance grows with smaller $p$.}
    \label{fig:bern}
\end{figure*}

\subsection{Graph sampling designs}\label{ssec:designs}

We consider a couple workhorse network sampling designs such as induced subgraph sampling (where nodes are first selected via Bernoulli sampling or simple random sampling without replacement) and traceroute sampling; see e.g.,~\cite{frank1980estimation, achlioptas2009bias}. Under Bernoulli sampling (BS), each node is retained in $\ccalV^*$ independently with probability \(p\). Induced subragph sampling dictates that an edge $(i,j)\in\ccalE$ is observed (and included in $\ccalE^*$) only when both $i,j\in\ccalV^*$. This yields the uniform edge inclusion probability \(\pi_{ij}=p^2\), for all \((i,j)\in\mathcal E^{*}\). Alternatively, in simple random sampling (SRS) one selects $n^*$ nodes uniformly at random and without replacement (as for testability in Section \ref{sec:testability}).  
Therefore, the node inclusion probability is $\pi_{i}=\frac{n^*}{n}$ and for induced subgraph sampling the edge inclusion probability is $\pi_{ij}=\frac{n^*(n^*-1)}{n(n-1)}$. Again, notice that inclusion probabilities do not depend on specific units. 

In contrast, traceroute sampling traverses and samples edges along shortest paths between \(n_{\mathrm S}\) randomly chosen sources and \(n_{\mathrm T}\) randomly chosen targets. Edges that appear frequently on shortest paths are more likely to be sampled, leading to non-uniform inclusion probabilities that depend on the topology of the graph. This is precisely an instance of unequal probability sampling. While it is challenging to derive the edge inclusion probabilities exactly, one can obtain the approximation
\begin{equation*}
\pi_{ij}\approx 1-\exp\!\left(-\,b_{ij}\,\frac{n_{\mathrm S}n_{\mathrm T}}{n^{2}}\right),
\end{equation*}
where \(b_{ij}\) denotes the betweeness centrality of edge $(i,j)$~\cite[Ch. 5]{kolaczyk_2009}. 
These three sampling designs illustrate the transition from uniform and readily obtained inclusion probabilities to highly heterogeneous ones, which can be intractable to compute -- a staple of network sampling.

In our ensuing numerical experiments, we will examine several of these canonical network sampling designs, as well as other homophily measures beyond $\textrm{TV}_G(\bbX)$~\cite[Sec. 3]{luan2024heterophily}. These analyses will demonstrate the robustness of the proposed estimator and reveal trade-offs related to the feasibility of evaluating inclusion probabilities and estimators' variances.

\begin{table*}[t]
\caption{Estimation of different homophily measures under SRS-based induced subgraph sampling. Ground truth (GT) values and HT estimates of Dirichlet energy, edge homophily and node homophily are shown for samples containing $30\%$ of the total node counts, averaged over $T=200$ realizations.}
\label{tab:iss_results}
\centering
\begin{tabular}{l rr rrr rrr rrr}
\toprule
& \multicolumn{2}{c}{Size}
& \multicolumn{3}{c}{Dirichlet energy}
& \multicolumn{3}{c}{Edge homophily}
& \multicolumn{3}{c}{Node homophily} \\
\cmidrule(lr){2-3}
\cmidrule(lr){4-6}
\cmidrule(lr){7-9}
\cmidrule(lr){10-12}
Dataset
& Nodes & Edges
& GT & Est & \cellcolor{gray!10}Bias
& GT & Est & \cellcolor{gray!10}Bias
& GT & Est & \cellcolor{gray!10}Bias \\
\midrule
Amazon
  & 334863 & 925872
  & 0.6196 & 0.6200 & \cellcolor{gray!10} 0.0004
  & 0.3804 & 0.3808 & \cellcolor{gray!10} 0.0005
  & 0.3757 & 0.3760 & \cellcolor{gray!10} 0.0003 \\
Citeseer
  & 3327 & 4676
  & 0.2575 & 0.2564 & \cellcolor{gray!10}-0.0010
  & 0.7425 & 0.7575 & \cellcolor{gray!10} 0.0150
  & 0.7102 & 0.7070 & \cellcolor{gray!10}-0.0032 \\
Cora
  & 2708 & 5429
  & 0.1900 & 0.1902 & \cellcolor{gray!10} 0.0002
  & 0.8100 & 0.8097 & \cellcolor{gray!10}-0.0002
  & 0.8252 & 0.8187 & \cellcolor{gray!10}-0.0064 \\
Cornell
  & 183 & 298
  & 0.8679 & 0.8817 & \cellcolor{gray!10} 0.0138
  & 0.1321 & 0.1698 & \cellcolor{gray!10} 0.0376
  & 0.1160 & 0.1411 & \cellcolor{gray!10} 0.0252 \\
Pubmed
  & 19717 & 44338
  & 0.1976 & 0.1981 & \cellcolor{gray!10} 0.0005
  & 0.8024 & 0.8021 & \cellcolor{gray!10}-0.0003
  & 0.7924 & 0.7971 & \cellcolor{gray!10} 0.0047 \\
Wisconsin
  & 251 & 450
  & 0.7940 & 0.7991 & \cellcolor{gray!10} 0.0052
  & 0.2060 & 0.2822 & \cellcolor{gray!10} 0.0762
  & 0.1665 & 0.2025 & \cellcolor{gray!10} 0.0360 \\
Question
  & 4897 & 15362
  & 0.1604 & 0.1590 & \cellcolor{gray!10}-0.0014
  & 0.8396 & 0.8410 & \cellcolor{gray!10} 0.0014
  & 0.8980 & 0.8887 & \cellcolor{gray!10}-0.0093 \\
Squirrel
  & 5201 & 217073
  & 0.7773 & 0.7775 & \cellcolor{gray!10} 0.0002
  & 0.2227 & 0.2239 & \cellcolor{gray!10} 0.0013
  & 0.2176 & 0.2189 & \cellcolor{gray!10} 0.0013 \\
Chameleon
  & 2277 & 31371
  & 0.7688 & 0.7633 & \cellcolor{gray!10}-0.0056
  & 0.2312 & 0.2332 & \cellcolor{gray!10} 0.0020
  & 0.2469 & 0.2442 & \cellcolor{gray!10}-0.0028 \\
Karate club
  & 34 & 78
  & 0.1082 & 0.1062 & \cellcolor{gray!10}-0.0013
  & 0.8918 & 0.8882 & \cellcolor{gray!10}-0.0036
  & 0.8882 & 0.8728 & \cellcolor{gray!10}-0.0154 \\
\bottomrule
\end{tabular}
\end{table*}

\section{Numerical Evaluation}\label{s:numerical_eval}
We assess the performance of the HT estimator for different homophily metrics using a variety of network datasets.
The code used to reproduce all the results we report is publicly available on GitHub\footnote{\url{https://github.com/hamedajorlou/Homophily-HT-Estimation}}, and the interested reader is referred therein for additional experiments and implementation details.

\vspace{2pt}\noindent \textbf{Experimental setup.}
Each graph considered in our experiments is equipped with categorical node attributes, represented as one-hot encoded vectors collected in $\bbX_n \in \mathbb{R}^{n \times f}$, where $f$ is the number of node-types. Given a graph $G$ and its node features $\bbX_n$, the task is to estimate the global homophily level from only a sampled subset of its edges.

To generate partial observations, we consider sampling schemes with both equal and unequal edge inclusion probabilities. Induced subgraph sampling (where nodes are selected using either BS or SRS) yields an equal--probability design, while traceroute sampling represents the unequal-probability regime. Each sampling method produces an observed subgraph $G^{*}$ with attributes $\bbX_{n^*}$, from which we compute an homophily estimate $\widehat{\textrm{TV}}_{\textrm{HT}}$ using the inclusion probabilities in Section \ref{ssec:designs}. Performance is evaluated over $T=200$ realizations by reporting the bias $(\widehat{\textrm{TV}}_{\textrm{HT}}-\textrm{TV}_G(\bbX_n))$. All Dirichlet energy values are normalized to the interval $[0,1]$.

\vspace{2pt}\noindent \textbf{Datasets and preliminary observations.}
Our experiments use the heterophily benchmark datasets introduced and standardized in recent evaluations of non-homophilous graphs~\cite{lim2021large, sun2022beyond, platonov2022critical, zhou2024opengsl}. The collection includes graphs such as Chameleon, Cora, Citeseer, Cornell, Amazon, Wisconsin, Squirrel, and Karate Club, which cover a wide range of network sizes and homophily levels. This diversity provides an ideal setting to assess homophily estimation under different sampling designs. Across all datasets, initial experiments confirm that the HT estimator recovers homophily metrics with negligible bias. We discuss these findings through several test cases presented next.

\vspace{2pt}\noindent \textbf{Test Case 1 - Dispersion analysis under different $p$.}  
In this experiment, we study how $p$ influences the variability of the HT estimator under BS sampling of nodes. We examine the sampling rates $p=\{0.1, 0.3, 0.5\}$, and recall the edge inclusion probability $\pi_{ij} = p^{2}$. Notice that $\widehat{\textrm{TV}}_{\textrm{HT}}=p^{-2}\widehat{\textrm{TV}}_{G^*}(\bbX_{n^*})$, so the HT estimator is a scaled-up correction of the plug-in estimator. As shown in the histograms in Fig.~\ref{fig:bern}, unbiasedness is supported for all values of $p$, but the dispersion changes significantly.  For small sampling rates, the estimator exhibits a wide spread due to the limited number of observed $V_{ij}$ values (even with the correction of $p^{-2}$). As the number of sampled edges grows, the distribution of estimates concentrates around the ground-truth value.

\vspace{2pt}\noindent \textbf{Test Case 2 - Different homophily measures.}
Here we evaluate the performance of the HT estimator under induced subgraph sampling. A fixed number of nodes is selected uniformly at random without replacement (SRS), and all edges between the selected nodes are observed. Table~\ref{tab:iss_results} reports the results obtained by sampling $30\%$ of the total nodes, averaged over $T=200$ realizations. Across all datasets, the estimator remains effectively unbiased: the estimated Dirichlet energy, edge homophily where $V_{ij}$ is replaced by $A_{ij}\ind{\bbx_j\equiv \bbx_j}$, and node homophily~\cite{luan2024heterophily} closely match their ground-truth values, with biases typically in the order of $10^{-3}$ or smaller. Even with a rather small sample size of edges, the consistency of the results across citation graphs, web networks, and smaller benchmark graphs confirms that the HT estimator performs reliably under induced subgraph sampling.

\vspace{2pt}\noindent \textbf{Test Case 3 - Traceroute sampling.} To close, we study the behavior of the HT estimator under traceroute sampling, a setting in which edges are discovered along shortest paths between randomly selected source and target nodes. 
We consider the Zachary Karate Club network using three distinct source--target pair sampling rates, to illustrate the estimator's behavior under different path configurations. Across all cases, the estimator remains unbiased, confirming that the weights correctly compensate for unequal inclusion probabilities. The recovered homophily metrics remain stable across all three source--target settings, corroborating that the HT estimator is capable of handling strongly heterogeneous sampling designs, provided that the inclusion probabilities are well approximated. 
\begin{figure}
    \includegraphics[width=0.47\textwidth]{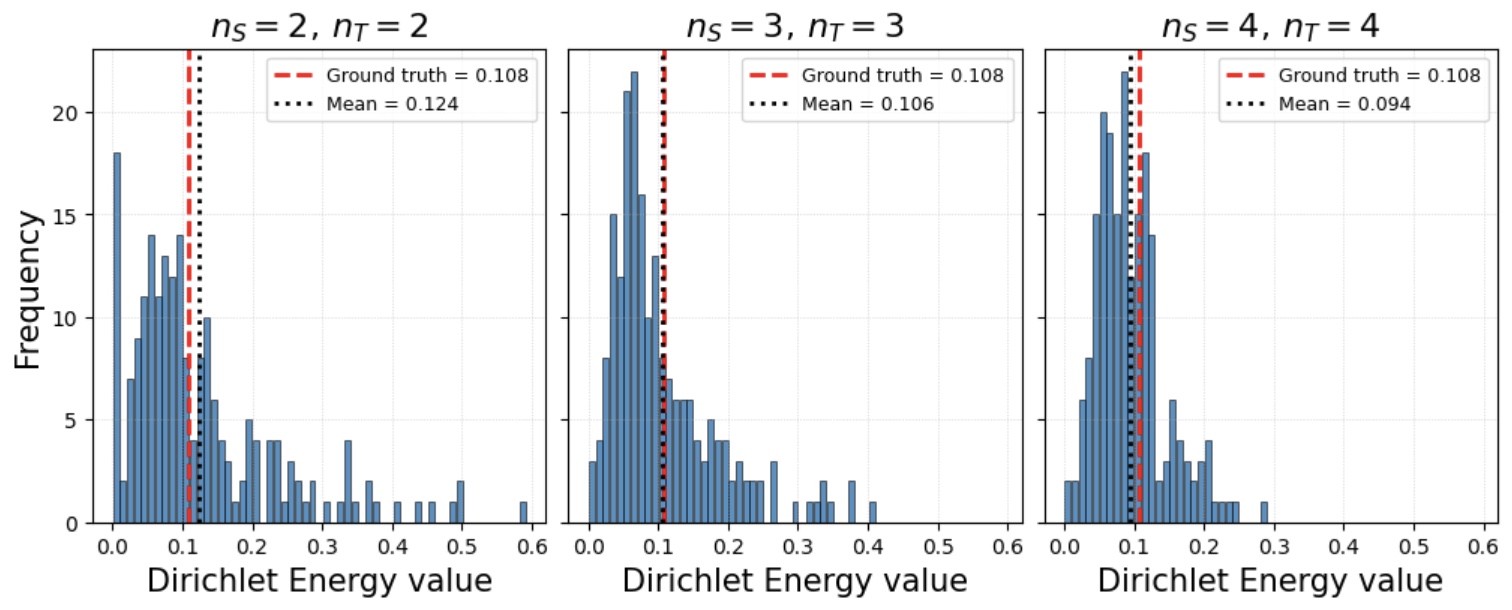}
    \caption{Homophily estimation under traceroute sampling. Shortest paths are traced between randomly selected source and target nodes on the Karate Club network. Histograms of the estimated Dirichlet energy for different source and target sampling rates, obtained for $T=200$ realizations.}
    \label{fig:trc}
\end{figure}

\section{Conclusions}\label{sec:conclusions}

This work introduced a unified framework for estimating homophily-related structural metrics from sampled graphs. Using the HT methodology, we derived unbiased estimators for the Dirichlet energy (and other heterophily measures expressible as edge totals or averages) under both unform and unequal-probability sampling designs. We also established that the Dirichlet energy is a testable (i.e., weakly consistent) graph parameter under induced subgraph sampling.

Our reproducible numerical experiments using benchmark graph datasets confirmed that the novel HT estimators remain unbiased when the population graphs are sampled via induced subgraph or traceroute sampling. These findings underscore the importance of accounting for the sampling design when inferring structural properties from partial network measurements, and the pitfalls of plug-in estimators. Future work includes developing variance-reduction strategies, extending the analysis to dynamic graphs, and examining testability for other heterophily-related network statistics.

\bibliographystyle{IEEEbib}
\bibliography{bibfile.bib}

\end{document}